\documentclass[pdflatex,sn-nature]{sn-jnl}% Style for submissions to Nature Portfolio journals

\usepackage{graphicx}%
\usepackage{multirow}%
\usepackage{amsmath,amssymb,amsfonts}%
\usepackage{amsthm}%
\usepackage{mathrsfs}%
\usepackage[title]{appendix}%
\usepackage{xcolor}%
\usepackage{textcomp}%
\usepackage{manyfoot}%
\usepackage{booktabs}%
\usepackage{algorithm}%
\usepackage{algorithmicx}%
\usepackage{algpseudocode}%
\usepackage{listings}%
\begin{document}

\title{Emergent dilemma and periodic oscillation in the nonlinear interplay between epidemic and behavior}


\author[1,4,7,8]{Longzhao Liu}

\author[9]{Hongwei Zheng}

\author[3]{Yajing Hao}

\author*[5]{Qun Wang}\email{qunwang@buaa.edu.cn}

\author*[1,4,7,8]{Xin Wang}\email{wangxin\_1993@buaa.edu.cn}

\author*[1,2,4,6,7,8]{Shaoting Tang}\email{tangshaoting@buaa.edu.cn}

\affil[1]{\small Institute of Artificial Intelligence, Beihang University, Beijing 100191, China}
\affil[2]{\small Hangzhou International Innovation Institute, Beihang University, Hangzhou 311115, China}
\affil[3]{\small School of Mathematical Sciences, Beihang University, Beijing, 100191, China}
\affil[4]{\small Key laboratory of Mathematics, Informatics and Behavioral Semantics, Beihang University, Beijing 100191, China}
\affil[5]{\small School of Logistics, Beijing Wuzi University, Beijing 101149, China}
\affil[6]{\small Institute of Medical Artificial Intelligence, Binzhou Medical University, Yantai 264003, China}
\affil[7]{\small Beijing Advanced Innovation Center for Future Blockchain and Privacy Computing, Beihang University, Beijing 100191, China}
\affil[8]{\small State Key Laboratory of Complex \& Critical Software Environment, Beihang University, Beijing 100191, China}
\affil[9]{\small Beijing Academy of Blockchain and Edge Computing, Beijing 100085, China}

\abstract{Human behaviors, particularly non-pharmaceutical interventions (NPIs), are dynamically coupled with epidemic spreading. While prior studies mainly assume a linear interplay, real-world behavioral evolution is driven by nonlinear responses and social influence. Here, we incorporate these multifaceted mechanisms into a co-evolutionary model and analytically derive the critical thresholds. Notably, as the infection rate grows, NPI compliance initially rises but then abruptly drops to zero. This paradoxical decline indicates an emergent social dilemma: at high infection rates, abandoning NPIs is individually optimal but detrimentally triggers an explosive surge in epidemic prevalence. We further show that socially induced overestimation of the infection rate can counterintuitively prompt individuals to abandon NPIs. Moreover, the interplay with social influence induces periodic oscillations, reflecting a tragic cycle of recurrent epidemic waves. Furthermore, we validate the robustness of this NPI-abandonment dilemma in networked population. Our work illustrates rich emergent phenomena in the co-evolution of epidemic and behavior, challenging traditional views on this coupled dynamics.}

\maketitle

\section{Introduction}
Epidemic spreading remains a major threat to global public health, and its trajectory is crucially influenced by human social behaviors \cite{bergstrom2024human, aleta2020modelling, funk2010modelling, oraby2014influence, liu2025higher}. For instance, during the COVID-19 pandemic, non-pharmaceutical interventions (NPIs) effectively reduced disease prevalence and saved countless lives \cite{howard2021evidence, flaxman2020estimating, perra2021non}. However, in the real world, individual compliance with such preventive measures is rarely guaranteed. Instead, it manifests as a dynamic process shaped by multiple factors, such as risk perception and social influence, which continuously co-evolves with epidemic dynamics itself \cite{saad2023dynamics,fenichel2011adaptive, chen2025simple}.  Consequently, unraveling the complex interplay between behavioral responses and epidemic spreading is critical for controlling infectious diseases.

Extensive research has explored the co-evolution of epidemic and behaviors, yielding important insights such as the dilemma of vaccination uptakes and the emergence of multiple thresholds \cite{gozzi2025comparative, moore2021vaccination, chen2019imperfect, st2024paradoxes, verelst2016behavioural, funk2009spread}. In particular, the Covid-19 pandemic demonstrates that NPIs may serve as the only effective countermeasures during early stages of newly emerging infectious diseases, thereby garnering increasing attention \cite{lai2020effect,block2020social,li2020substantial}. A key characteristic of NPIs, such as mask wearing and social distancing, is their transient nature: they offer protection only during active compliance, with their efficacy ending immediately upon cessation \cite{eikenberry2020mask, stockmaier2021infectious}. This implies that the dynamics of NPI usage and epidemic transmission interact on a comparable timescale \cite{glaubitz2024social}. To model their interplay, researchers typically utilize contagion frameworks and evolutionary game theory \cite{traulsen2023individual,jiang2026nonlinear}. Specifically, the latter frames NPI adoption as a strategic decision where individuals weigh socioeconomic costs of compliance against the benefits of mitigating infection risks \cite{nicola2020socio,reluga2010game}. These coupled models have revealed non-trivial phenomenon absent in isolated epidemic dynamics. For instance, there emerges the oscillating tragedy where multiple infection peaks appear \cite{glaubitz2020oscillatory, weitz2020awareness, ye2021game, granell2013dynamical}. Moreover, a higher infection rate counterintuitively leads to a smaller final epidemic size, as it elevates public risk perception, driving broader NPI adoption \cite{morsky2023impact, noori2025coevolution, qiu2022understanding}. This insight suggests that amplifying perceived risk of being infected, through information dissemination, is always beneficial for epidemic control.

\begin{figure}[tb]
\centering
\includegraphics[width=.9\linewidth]{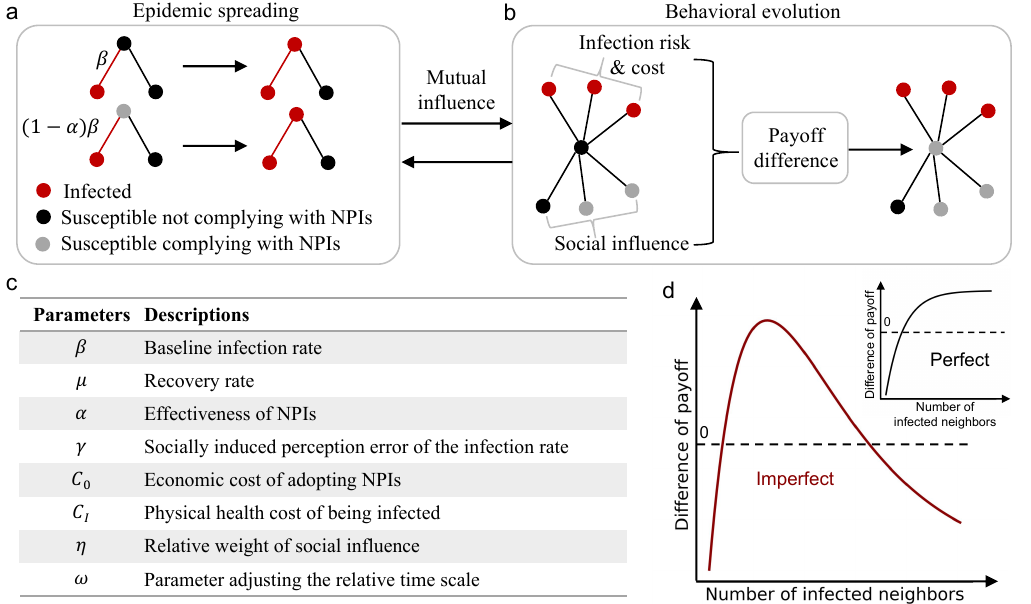}
\caption{Co-evolutionary model of epidemic and behavior. The disease transmission dynamics incorporate the effectiveness of NPIs (a), while behavioral evolution is driven by nonlinear perceived infection risk and social influence (b). Detailed parameters and descriptions are shown in (c). (d) The payoff difference between NPI compliance and non-compliance is plotted against the number of infected contacts under imperfect protection (main panel) and perfect protection (inset). We find the non-monotonic change of payoff difference.}
\label{fig1}
\end{figure}

Despite the progress, most works assume linear behavioral responses to epidemic risks, whereas human decision-making is inherently nonlinear. Although rare exceptions consider nonlinear mechanisms, their assumption of perfect NPIs precludes the fundamental non-monotonicity in payoff differences, which is driven by the combined effect of nonlinearity and imperfect protection (see Fig.\ref{fig1}d) \cite{glaubitz2020oscillatory}. Furthermore, beyond individual risk assessment, social influences including social norms and peer pressure serve as fundamental drivers of behavioral evolution \cite{liu2020homogeneity, watts2002simple, baxter2022local, morsky2023impact, weston2018infection}. Yet, how these multifaceted mechanisms shape macroscopic emergent dynamics remains an open question.

In this work, we propose an epidemic-behavior co-evolutionary model that incorporates both nonlinear responses to risks and social influences, and theoretically analyze critical behaviors. Notably, the inherent nonlinearity induces rich emergent phenomena including bistability and a counterintuitive dilemma. Specifically, as the infection rate increases, NPI adoption initially rises but then abruptly drops to zero. This decline challenges the intuition that greater infection risks promote NPI adoption, highlighting an emergent dilemma: at high infection rates, abandoning NPIs is individually optimal but collectively detrimental, ultimately exacerbating the epidemic. In particular, socially induced overestimation of infection risk may facilitate the emergence of the dilemma. Furthermore, when coupled with social influence, the system behavior can transit from a stationary prevalence to periodic oscillations. Moreover, the nonlinearity-induced phenomena similarly persist on structured populations. Our work advances the understanding of epidemic-behavior systems, transcending traditional insights gleaned from linear assumptions.

\section{Model framework}
We consider the joint dynamics of disease spreading and behavioral evolution, as shown in Fig.\ref{fig1}. The disease dynamics are characterized by the classic susceptible-infected-susceptible (SIS) model. Specifically, infected individuals transmit the disease to unprotected susceptible ones at a baseline rate of $\beta$, whereas this rate is reduced to $(1-\alpha)\beta$ if susceptible individuals adopt NPIs. Here, $\alpha$ measures the effectiveness of NPIs: $\alpha=1$ implies perfect protection, while $0<\alpha<1$ corresponds to more realistic imperfect protection. Moreover, infected individuals become susceptible again at a rate $\mu$, and simultaneously revert to their pre-infection NPI status due to behavioral inertia. 

Then, we utilize evolutionary game framework to model the behavioral evolution which is driven by multiple factors (see Fig.\ref{fig1}b). First, adopting NPIs, such as mask-wearing and social distancing, incurs personal discomfort and economic losses, which are depicted by a cost parameter $C_0$. Second, the behavioral decision-making relies on individual perceived infection risk and cost. Let $I(t)$ denote the density of infected individuals. For susceptible individuals without NPI protection, their perceived risk can be approximated by $(1-e^{-\gamma\beta k I(t)})$. Thus, their perceived infection cost, denoted by $\pi_{n,I}$, satisfies 
\begin{equation}
\pi_{n,I} = -C_I (1-e^{-\gamma\beta k I(t)})
\end{equation}
$C_I$ is the physical health cost, and $k$ is the number of contacts per unit time. The parameter $\gamma$ depicts the perception error of the infection rate, which is mainly influenced by social factors such as information diffusion. Specifically, $\gamma>1$ means socially induced overestimate, while $\gamma<1$ implies an underestimation. Similarly, for susceptible individuals complying with NPIs, their perceived infection cost can be written as
\begin{equation}
\pi_{c,I} = -C_I (1-e^{-(1-\alpha)\gamma\beta k I(t)})
\end{equation}
In addition, we incorporate social influence and quantify its payoff based on the principle that it is positively associated with the fraction of individuals adopting similar behaviors. Without loss of generality, we use the simplest functional form: the social payoff for NPI compliance is given by  $\pi_{c,s}=\eta (p(t)-(1-p(t)))$, while the social payoff for non-compliance is $\pi_{n,s}=\eta ((1-p(t))-p(t))$. Here, $p(t)$ denotes the fraction of compliers in the susceptible population, and $\eta$ measures the relative weight of social influence.

Overall, the total payoffs of NPI compliance and non-compliance can be respectively written as
\begin{equation}
\begin{aligned}
\pi_{c} &= -C_0 -C_I (1-e^{-(1-\alpha)\gamma\beta k I(t)}) + \eta (2p(t)-1)\\
\pi_{n} &= -C_I (1-e^{-\gamma\beta k I(t)}) + \eta (1-2p(t))
\end{aligned}
\end{equation}
The payoff difference of these two behaviors is denoted by $\Delta\pi = \pi_{c}-\pi_{n}$. Notably, the payoff difference varies non-monotonically with $I(t)$, as shown in Fig.\ref{fig1}d, in contrast to the pattern observed under the assumptions of linear perceived risk and perfect protection.

Finally, according to contagion framework and replicator dynamics, we derive the governing equations of this co-evolutionary model:
\begin{equation}
\begin{aligned}
\frac{dI}{dt} &= -\mu I + \beta k (1-\alpha p)I(1-I)\\
\frac{dp}{dt} &= \omega p (1-p) \{-C_0 -C_I (e^{-\gamma\beta k I}-e^{-(1-\alpha)\gamma\beta k I}) + 2\eta (2p-1)\}
\end{aligned}
\label{evoeq}
\end{equation}
where $\omega$ adjusts the relative time scale.

\section{Results}

\begin{figure*}[tbhp]
\centering
\includegraphics[width=.95\linewidth]{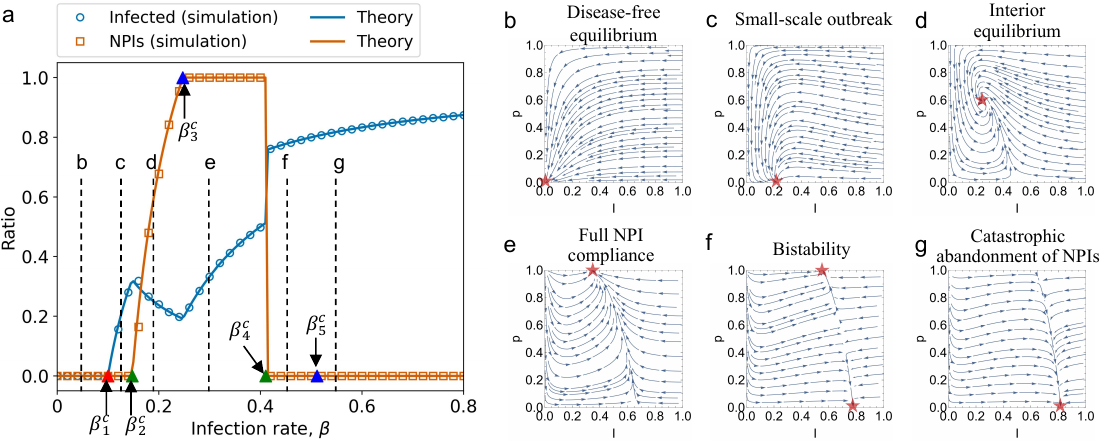}
\caption{Catastrophic abandonment of NPIs and bistability. (a) We present the fraction of infected individuals and NPI compliance as a function of $\beta$. They exhibit complex changes with multiple phase transitions. An intriguing transition is the collective abandonment of NPIs, which induces explosive growth in epidemic prevalence. The triangles represent the thresholds $\beta_2^c$--$\beta_5^c$, which are solved by Eq.\eqref{eq5} and Eq.\eqref{eq6}. (b)-(g) Vector fields are shown under six representative $\beta$. They correspond to six phases: disease-free equilibrium, small-scale outbreak without NPI compliance, stable interior equilibrium, full NPI compliance, bistability and catastrophic abandonment of NPIs. Parameters: $N=2000, k=10, \alpha=0.5, \mu=1, \gamma=1, C_0=1, C_I=6, \eta=0.001, \omega=1$.}
\label{fig2}
\end{figure*}
\subsection{Emergent dilemma and bistability induced by the nonlinearity}
We first perform both theoretical analysis and model simulations in well-mixed populations (see simulation details in Supplementary Information), to explore the impact of nonlinear behavioral responses. Fig.\ref{fig2}a shows the complex changes in epidemic prevalence and NPI compliance level, exhibiting multiple phase transitions. Specifically, as infection rate $\beta$ grows, NPI compliance would increase, thereby reducing epidemic prevalence, which confirms the classic finding from prior works \cite{noori2025coevolution, qiu2022understanding}. Yet, as $\beta$ grows further beyond a critical value, the fraction of NPI compliance surprisingly abruptly drops to zero, thereby triggering an explosive increase in epidemic prevalence. This finding challenges conventional wisdom that greater infection risks enhances the willingness to adopt NPIs. More critically, this finding unveils an emergent dilemma: when $\beta$ is high, abandoning NPIs is individually optimal, while it severely exacerbates epidemic prevalence across the population. 

This counterintuitive phenomenon can be naturally explained by a microscopic mechanism: the nonlinear perceived risk. As shown in Fig.\ref{fig1}d, the nonlinearity induces non-monotonic changes in payoff difference between NPI compliance and non-compliance. Specifically, once the fraction of infected individuals exceeds a value, the payoff of NPI compliance falls below that of non-compliance. This drives individuals to abandon NPIs, thereby leading to higher infection prevalence, which in turn enlarge the relative advantage of non-compliance. Such a vicious feedback loop induces the explosive decrease in NPI compliance.

Moreover, our analysis reveals that the threshold for the dilemma is sensitive to initial conditions (see Fig.S1 in Supplementary Information). Specifically, a higher initial density of NPI compliance can postpone this critical threshold. This finding gives rise to another intriguing phenomenon, i.e., a bistable regime, where final state could be effectively controlled by modulating initial conditions.

To give a theoretical insight, we analytically derive threshold formulas for these phase transitions, through exploring stability of different equilibriums $(I^*, p^*)$ in Eq.\eqref{evoeq} (see Supplementary Information Sec.S2). Specifically, $\beta_1^c = \mu/k$ represents the epidemic invasion threshold, derived from the stability condition of the disease-free equilibrium $(0,0)$. Furthermore, through analyzing the equilibrium $(1-\mu/\beta k,0)$, we obtain critical conditions for both the emergence and the abrupt disappearance of NPI compliance. These critical values, indicated by $\beta_2^c$ and $\beta_4^c$ in Fig.\ref{fig2}a, are the two roots of the following equation:
\begin{equation}
e^{-(1-\alpha)\gamma(\beta k - \mu)} - e^{-\gamma(\beta k - \mu)} = \frac{2\eta+C_0}{C_I}
\label{eq5}
\end{equation}
Then, we analyze the case under large initial density of NPI compliance, which is associated with the equilibrium $(1-\mu/(1-\alpha)\beta k,1)$. Its stability condition corresponds to the lower and upper thresholds for full NPI compliance, indicated by $\beta_3^c$ and $\beta_5^c$ in Fig.\ref{fig2}a, which can be computed by
\begin{equation}
e^{-(1-\alpha)\gamma(\beta k - \frac{\mu}{1-\alpha})} - e^{-\gamma(\beta k - \frac{\mu}{1-\alpha})} = \frac{C_0-2\eta}{C_I}
\label{eq6}
\end{equation}
The region enclosed by $\beta_4^c$ and $\beta_5^c$ constitutes the bistable regime. Fig.\ref{fig2}a demonstrates that our theory accurately predicts the final prevalence and critical behaviors observed in large-scale simulations.

These critical thresholds divide the system into six phases, with their representative vector fields presented in Fig.\ref{fig2}b-g. Notably, Fig.\ref{fig2}f shows critical initial conditions within the bistable regime, which determine whether the entire susceptible population adopts or abandons NPIs. Fig.\ref{fig2}g illustrates the collective abandonment of NPIs at large $\beta$. 

\begin{figure*}[bt]
\centering
\includegraphics[width=.95\linewidth]{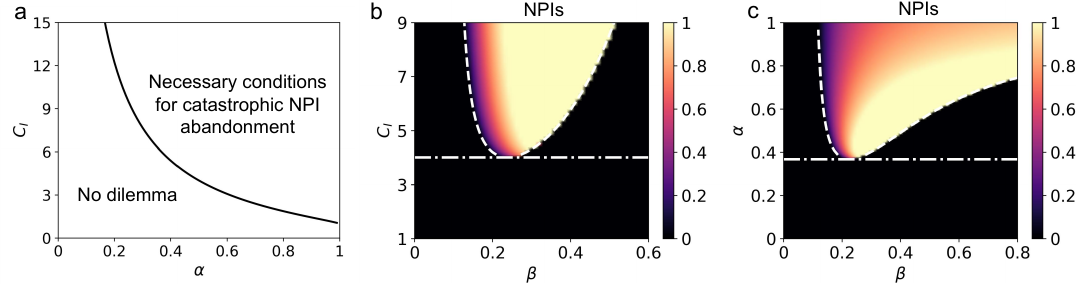}
\caption{Critical conditions for the catastrophic NPI abandonment. (a) The black line is computed by Eq.\eqref{eq7}, which represents the necessary conditions for the dilemma to emerge. (b)--(c) Phase diagrams illustrating the ratio of NPI compliance are shown under (b) $\alpha=0.5$ and (c) $C_I=6$. The dash-dotted line is calculated by Eq.\eqref{eq7} and the dashed lines are numerically solved by Eq.\eqref{eq5}, which all accurately predict the thresholds. The results reveal that the dilemma about catastrophic NPI abandonment stably exists in a wide parameter space. Parameters: $N=2000, k=10, \mu=1, \gamma=1, C_0=1, \eta=0.001, \omega=1$.}
\label{fig3}
\end{figure*}
We further examine the conditions for the emergent dilemma, i.e., catastrophic abandonment of NPIs. For a small initial fraction of NPI compliance, we note that this phenomenon occurs if and only if Eq.\eqref{eq5} yields two solutions for $\beta$. When $\alpha=1$, the equation has at most one solution for $\beta$, demonstrating that perfect protection could not induce the catastrophic abandonment of NPIs. When $0<\alpha<1$, by comparing the maximum of the left-hand side with the right-hand side of Eq.\eqref{eq5}, we analytically derive the condition for the existence of two solutions, which is
\begin{equation}
C_I > \frac{(1-\alpha)(2\eta + C_0)}{\alpha}* e^{-\frac{\ln(1-\alpha)}{\alpha}}
\label{eq7}
\end{equation}
Eq.\eqref{eq7} reveals the relation between $C_I$ and $\alpha$, which serves as the necessary condition for the dilemma to emerge. This condition is represented by the critical line in Fig.\ref{fig3}a. More specifically, Fig.\ref{fig3}b-c present phase diagrams illustrating the fraction of NPI compliance under joint effect of $C_I$ and $\beta$ as well as joint effect of $\alpha$ and $\beta$, where the white lines represent the thresholds. Phase diagrams of epidemic prevalence are shown in Supplementary Information. As illustrated by dash-dotted lines in Fig.\ref{fig3}b-c, the results validate the predictive accuracy of Eq.\eqref{eq7}  and demonstrate that the emergent dilemma exists within a wide range of parameters. Furthermore, we find that the threshold for NPI abandonment increases with $C_I$ and $\alpha$.

\begin{figure*}[tb]
\centering
\includegraphics[width=.95\linewidth]{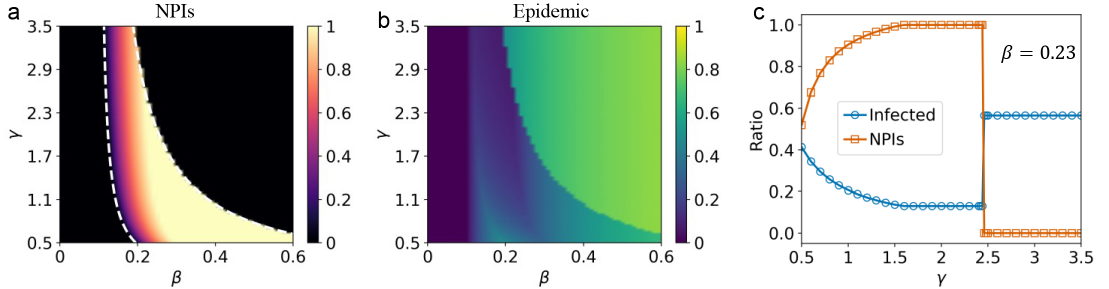}
\caption{Impact of socially induced perception error ($\gamma$). (a)--(b) Phase diagrams about (a) ratio of NPI compliance and (b) ratio of infected individuals are shown under the joint effect of $\beta$ and $\gamma$. The white lines, numerically solved by Eq.\eqref{eq5}, depict critical thresholds. Results illustrate that $\gamma$ has a two-fold impact on disease spreading. (c) The final state is plotted against $\gamma$ for a representative value of $\beta=0.23$. We find that epidemic prevalence first decreases and then explosively grows as $\gamma$ increases. Parameters: $N=2000, k=10, \alpha=0.5, \mu=1, C_0=1, C_I=6, \eta=0.001, \omega=1$.}
\label{fig4}
\end{figure*}
In Fig.\ref{fig4}, we explore the perception error regarding the infection rate, characterized by $\gamma$, which is sensitive to social factors including information dissemination and public awareness campaigns. For instance, intensive media coverage may lead individuals to overestimate the infection rate, which is captured by $\gamma>1$. Fig.\ref{fig4}a-b demonstrate that socially induced over-perception exerts a two-fold impact. On the positive side, it reduces the emergence threshold for NPI compliance, thereby inhibiting epidemic spreading. Conversely, it lowers the threshold for NPI abandonment, which means that the explosive growth of the disease occurs at smaller $\beta$. Results further show that whether the overall impact of perception error is positive or negative depends heavily on disease status. Specifically, for a small $\beta$ (see Fig.\ref{fig4}c), moderate over-perception enhances NPI compliance and reduces epidemic size, whereas excessive overestimation causes collective NPI abandonment, leading to rapid growth of diseases. For a large $\beta$, under-perception ($\gamma<1$) actually proves more effective for disease control (see Fig.\ref{fig4}b). These findings imply that public health messaging must adapt to the disease status: overly aggressive campaigns are not universally beneficial and can even be counterproductive, particularly when the actual infection rate is high.

\subsection{Periodic oscillations upon incorporating social influence}

\begin{figure*}[tb]
\centering
\includegraphics[width=.95\linewidth]{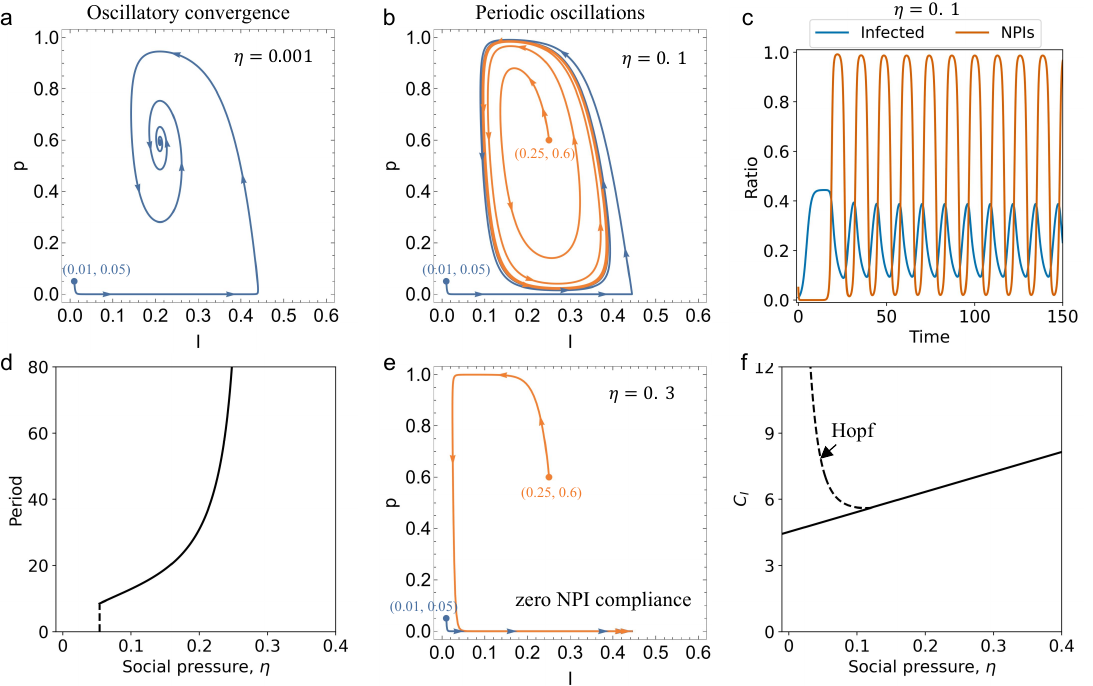}
\caption{The emergence of periodic oscillations after incorporating social influence. (a)--(b) The trajectories are plotted for (a) $\eta=0.001$ and (b) $\eta=0.1$. Each color corresponds to a distinct initial condition. At $\eta=1$, a limit cycle occurs, indicating periodic oscillations. The corresponding temporal evolution is shown in panel (c). (d) The oscillation periods are plotted against $\eta$. (e) For $\eta=0.3$, the trajectories again converge to a stable equilibrium. (f) Critical thresholds of $\eta$ for varying $C_I$. The dashed line, computed numerically using MatCont \cite{dhooge2008new}, marks the Hopf bifurcation where the system transitions from a stable equilibrium to periodic oscillations. The solid line, derived from Eq.\eqref{eq5}, represents the threshold at which NPI compliance vanishes. Parameters: $k=10, \alpha=0.5, \beta=0.18, \mu=1, \gamma=1,  C_0=1, C_I=7, \omega=5$.}
\label{fig5}
\end{figure*}

We extend our analysis by incorporating social influence, where its relative weight is denoted by $\eta$. For small $\eta=0.001$ (see Fig.\ref{fig5}a), the temporal evolution of the system initially undergoes oscillations and eventually reaches an interior equilibrium. This corresponds to a steady state where a fraction of individuals complies with NPIs. A marginal increase in $\eta$ enhances NPI compliance level, reducing disease prevalence (see Fig.S3 in Supplementary Information). Notably, as $\eta$ continues to increase, the system undergoes a Hopf bifurcation, where its final state shifts from a stable equilibrium to a limit cycle. More specifically, Fig.\ref{fig5}b presents a representative case of $\eta=0.1$. According to the Poincar\'e-Bendixson theorem, the trajectories originating from distinct initial conditions converge to a limit cycle. This corresponds to periodic oscillations in both NPI compliance and epidemic prevalence, as shown in Fig.\ref{fig5}c. This phenomenon implies a tragedy of public health that the population periodically experiences the infection peaks.

Furthermore, as evidenced by the case of $\eta=0.2$ (see Fig.S3 in Supplementary Information), the periodic oscillations are sustained across a range of $\eta$. Fig.\ref{fig5}d demonstrates that their periods increase rapidly as $\eta$ grows. This period expansion is because the limit cycle moves closer to the saddle equilibrium $(1-\mu/\beta k, 0)$, where the trajectory experiences a drastically reduced evolutionary speed. This also implies a prolonged infection peak within each cycle. However, as $\eta$ grows further beyond the threshold derived by Eq.\eqref{eq5}, the equilibrium $(1-\mu/\beta k, 0)$ becomes stable and the limit cycle vanishes (see Fig.\ref{fig5}e). In this situation, the population faces a worst-case scenario where no one complies with NPIs, thereby leading to a widespread disease outbreak. 
Fig.\ref{fig5}f illustrates these two thresholds of $\eta$ across varying infection cost $C_I$, which are indicated by the black lines. As $C_I$ grows, the critical value of $\eta$ inducing the periodic oscillations decreases, while the threshold leading to zero NPI compliance increases.

\subsection{Simulations in networked populations}

\begin{figure*}[tb]
\centering
\includegraphics[width=.95\linewidth]{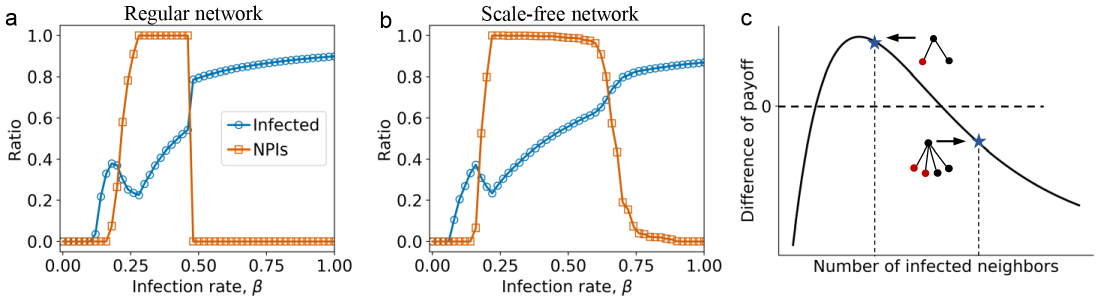}
\caption{Results in networked populations. We present simulation results on (a) a regular network and (b) a heterogeneous scale-free network generated by Barab\'asi-Albert algorithm \cite{barabasi1999emergence}. Both show that NPI compliance level exhibits a non-monotonic trend as $\beta$ grows. The main discrepancy is that the decline occurs more gradually on the scale-free network. (c) An intuitive explanation for this discrepancy. The payoff difference between abandoning and adopting NPIs is plotted against the number of infected neighbors. As indicated by the star markers, the payoff differences for a 2-degree node and a 4-degree node exhibit opposite signs, prompting the nodes to adopt divergent strategies. Parameters: $N=2000, \langle k \rangle=10, \alpha=0.5, \mu=1, \gamma=1, C_0=1, C_I=6, \omega=1, \eta=0.001$.}
\label{fig6}
\end{figure*}

We further consider network structure of populations and perform model simulations by using Gillespie algorithm (see details in Supplementary Information). First, Fig.\ref{fig6}a shows the case on a regular network where each node has the same degree. Similarly, as the infection rate $\beta$ grows, the NPI compliance level initially increases but then abruptly drops to zero. This indicates that the catastrophic NPI abandonment, i.e., the emergent dilemma, persists on regular networks. 

Fig.\ref{fig6}b shows the results on heterogeneous networks. We find that the pattern of an initial rise followed by a decline in NPI compliance remains valid. In contrast, this decline is more gradual than that on regular networks. This discrepancy arises because network heterogeneity induces a divergence in optimal strategies of individuals. Specifically, for a given infection prevalence, abandoning NPIs yields a higher payoff for higher-degree nodes due to their increased contact with infected neighbors, while maintaining NPIs is more advantageous for lower-degree nodes, as indicated by stars in Fig.\ref{fig6}c. 

Overall, the decline in NPI compliance persists across networked populations, underscoring the robustness of this dilemma.

\section{Conclusion}
Human behavioral evolution and epidemic spreading are deeply intertwined in the real world. However, unraveling this interplay and its ultimate impact remains a challenge, largely due to the inherent complexity introduced by nonlinear behavioral responses and social influences. This work proposes a co-evolutionary model by incorporating these multifaceted mechanisms and theoretically analyzes the critical thresholds. Crucially, we demonstrate that these mechanisms give rise to rich emergent phenomena. 

First, as the infection rate increases, NPI compliance initially rises, resulting in a non-monotonic trend in epidemic prevalence, which confirms the classic finding that greater infectivity can lead to a smaller epidemic \cite{noori2025coevolution, qiu2022understanding}. A fundamentally distinct finding is that once the infection rate surpasses a critical threshold, the NPI compliance level abruptly drops to zero, thereby triggering an explosive rise in epidemic prevalence. This abrupt decline indicates a social dilemma: abandoning NPIs becomes individually optimal at high infection rates, yet this behavior is collectively detrimental. Furthermore, we reveal that the threshold for this abrupt decline is sensitive to initial conditions, indicating the presence of bistability. Mechanistically, these findings are ultimately rooted in the nonlinear behavioral responses. 

Moreover, we investigate perception error for the infection rate driven by social factors including information diffusion. Notably, socially induced overestimation promotes NPI compliance and suppresses epidemic prevalence at low actual infection rates, but produces the opposite impact at high rates. This challenges the conventional understanding that higher perceived risk invariably translates to wider NPI adoption, implying that communicating high infection risks through social campaigns is not always effective.

Another emergent phenomenon is the periodic oscillation arising from social influences. Under weak social influences, the system exhibits oscillatory convergence characterized by multiple infection peaks, successfully reproducing the classic "oscillatory tragedy" \cite{glaubitz2020oscillatory, weitz2020awareness}. Going beyond this classical baseline, our analysis reveals that as social influence intensifies, the system undergoes a Hopf bifurcation and forms a limit cycle. This transition shifts the dynamics into sustained periodic oscillations, trapping the population in a vicious cycle marked by endless epidemic waves.

Furthermore, this dilemma, manifested as NPI abandonment at high infection rates, remains robust in networked populations.  We further show that individuals' optimal behaviors vary with their degrees on heterogeneous networks. This insight may inspire a 'divide and conquer' intervention strategy that targets specific degree groups lacking motivation to adopt NPIs.  

Our work unveils the complex phenomena induced by nonlinear behavioral responses and social influences, challenging the conventional understanding of epidemic-behavior dynamics. Despite the insights, several limitations warrant further investigation. For instance, although our current analysis focuses on the classic SIS model, the framework and co-evolutionary mechanisms can be extended to other epidemic models \cite{pastor2015epidemic}. Additionally, a rigorous theoretical framework is required to mathematically prove the existence of periodic behaviors within complex networks. Moreover, incorporating additional behavioral features, such as bounded rationality and inertia, deserves further exploration \cite{rubinstein1998modeling}.

\backmatter

\bmhead{Data availability}
There are no empirical data associated with this study. The simulated data supporting the findings of this study can be reproduced using the code provided.

\bmhead{Code availability}
The code is available at \url{https://github.com/llz154/Epidemic-behavior-dynamics}

\bmhead{Acknowledgements}
This work is supported by National Science and Technology Major Project (2022ZD0116800), Program of National Natural Science Foundation of China (12425114, 12201026, 12501702, 62441617), the Fundamental Research Funds for the Central Universities, Beijing Natural Science Foundation (Z230001), the Opening Project of the State Key Laboratory of General Artificial Intelligence (SKLAGI2025OP16) and Bejing Advanced Innovation Center for Future Blockchain and Privacy Computing.

\bmhead{Author contribution} L.L., Q.W., X.W. and S.T. conceived and designed the study. L.L., H.Z. and Y.H. conducted the simulations and theoretical analysis. L.L. wrote the initial draft, and all authors contributed to reviewing and editing the final manuscript.

\bmhead{Competing interests} The authors declare no competing interests.

\bmhead{Supplementary information} Supplementary information is available for this paper.


\end{document}